\documentclass[10pt, journal]{IEEEtran}

\usepackage{booktabs}
\usepackage{threeparttable}
\usepackage{amsmath,graphicx,cite,amssymb}
\usepackage{amsfonts,multirow,bm,array,setspace,stfloats}

\usepackage{graphicx,float,cite,amssymb}
\usepackage{graphics}
\DeclareGraphicsExtensions{.pdf,.jpeg,.png,.jpg}   
\usepackage{multirow,bm,bbm,array,setspace}
\usepackage{textcomp}
\usepackage{psfrag}
\usepackage{color}
\usepackage{pstricks,enumerate}
\usepackage{bbm}
\usepackage{makecell}
\usepackage{subfigure}
\usepackage{xpatch}
\usepackage{algorithm,algpseudocode}
\makeatletter
\newcommand{\distas}[1]{\mathbin{\overset{#1}{\kern\z@\sim}}}%
\newsavebox{\mybox}\newsavebox{\mysim}
\newcommand{\distras}[1]{%
	\savebox{\mybox}{\hbox{\kern1pt$\scriptstyle#1$\kern1pt}}%
	\savebox{\mysim}{\hbox{$\sim$}}%
	\mathbin{\overset{#1}{\kern\z@\resizebox{\wd\mybox}{\ht\mysim}{$\sim$}}}%
}

\ExplSyntaxOn
\cs_new:Npn \bibColoredItems #1#2
  {
    \clist_map_inline:nn {#2} { \cs_new:cpn {bib@colored@##1} {#1} } 
  }
\ExplSyntaxOff

\newcommand\bib@setcolor[1]{%
  \ifcsname bib@colored@#1\endcsname
    \expandafter\color\expandafter{\csname bib@colored@#1\endcsname}
  \else
    \normalcolor
  \fi
}

\xpatchcmd\@bibitem
  {\item}
  {\bib@setcolor{#1}\item}
  {}{\PatchFailed}

\xpatchcmd\@lbibitem
  {\item}
  {\bib@setcolor{#2}\item}
  {}{\PatchFailed}

\makeatother
\ifCLASSINFOpdf
\else
\fi

\newtheorem{proposition}{Proposition}

\newcommand{\bA}{\bm A}

\newcommand{\bG}{\bm G}

\newcommand{\bs}{\bm{s}}
\newcommand{\bI}{\bm{I}}

\newcommand{\by}{\bm{y}}

\newcommand{\bF}{\bm{F}}
\newcommand{\bH}{\bm{H}}

\newcommand{\bW}{\bm{W}}
\newcommand{\bD}{\bm{D}}

\newcommand{\hh}{\mathrm{H}}

\allowdisplaybreaks[4]

\begin{document}
	%
\title{Blind Passive Beamforming for MIMO System}
\author{
\IEEEauthorblockN{
    Wenhai Lai, \IEEEmembership{Graduate Student Member,~IEEE}, Jiawei Yao, and Kaiming Shen, \IEEEmembership{Senior Member,~IEEE}
} 
\thanks{
Accepted to IEEE Wireless Communications Letters on \today. \emph{(Corresponding author: Kaiming Shen.)}

The authors are with The Chinese University of Hong Kong, Shenzhen, China (e-mails: wenhailai@link.cuhk.edu.cn; jiaweiyao@link.cuhk.edu.cn; shenkaiming@cuhk.edu.cn).
}
}

%


\maketitle

\begin{abstract}
Passive beamforming for the intelligent surface (IS)-aided multiple-input multiple-output (MIMO) communication is a difficult nonconvex problem. It becomes even more challenging under the practical discrete constraints on phase shifts. Unlike most of the existing approaches that rely on the channel state information (CSI), this work advocates a blind beamforming strategy without any CSI. Simply put, we propose a statistical method that learns the main feature of the wireless environment from the random samples of received signal power. Field tests in the 5G commercial network demonstrate the superiority of the proposed blind passive beamforming method.
\end{abstract}


\section{Introduction}

Intelligent surface (IS) \cite{wu2024intelligent}, which is also known as reconfigurable intelligent surface (RIS) or passive antennas \cite{li2019towards, Arun2020RFocus},
comprises an array of reflective elements (REs), each giving rise to a reflected path from transmitter to receiver. A common purpose of IS is to modify the wireless environment by coordinating the phase shifts across the REs, namely the passive beamforming, in order to enhance the multiple-input multiple-output (MIMO) transmission. Unlike most relevant works in the literature that rely on channel state information (CSI), this work proposes a completely data-driven method, by which the active beamforming at transmitter requires low-dimensional CSI while the passive beamforming at IS does not require any CSI, namely \emph{blind passive beamforming}.

The existing works on the passive beamforming problem typically adopt the model-driven approach of first estimating the channels and then optimizing the phase shift of all REs, e.g., \cite{sun2024power} uses a single-layer neural network to acquire the CSI. However, channel estimation faces some formidable challenges in engineering practice, mainly recognized in the following two respects. First, each reflected signal is weak and can be easily overwhelmed by other signals and background noise. Second, the additional channel estimation for the IS is incompatible with the current network protocol because the IS then requires extra overhead and also the access to the communication chip of the user terminal.
As such, even though the channel estimation for IS has been extensively studied in the literature to date, the existing prototype realizations of IS \cite{Arun2020RFocus,pei2021ris, ren2022configuring, Xu2024Coordinating} seldom consider the channel estimation.

The second main difficulty of the model-driven approach lies in the optimization aspect, even after the CSI has somehow been perfectly obtained. Extensive efforts can be found in this area. The authors of \cite{MIMO_Zhangsw} suggest two different approximations of the MIMO capacity maximization problem in the low signal-to-noise ratio (SNR) regime and the high SNR regime, based on which the phase shift variables are optimized sequentially across the REs. The above method is extended to a double-IS system in \cite{2022Han_doubleIRS}. For the same MIMO capacity maximization problem, \cite{2022BaTWC} advocates the use of the Riemannian conjugate gradient method. Moreover, a variety of other standard methods including the semidefinite relaxation (SDR) \cite{2023LiJISAC}, the quadratic transform for fractional programming (FP) \cite{2022SabaTIFS}, the successive convex approximation (SCA) \cite{2022WzrMassiveMIMO}, and the majorization-minimization (MM) \cite{jiang2021joint} can be found in the literature. All the above methods require high computational complexity. In this paper, we show that the MIMO channel capacity maximization problem can be approximated as a more tractable sum power maximization problem, giving rise to a search-based method with linear complexity. More importantly, we show that the search-based method can even be realized without any CSI.

Our work is most closely related to the existing methods in \cite{Arun2020RFocus,ren2022configuring} based on the conditional sample mean (CSM) \cite{feller1968prob} of received signal power. But our work can be distinguished from \cite{Arun2020RFocus,ren2022configuring} in the following three respects. First,  \cite{Arun2020RFocus,ren2022configuring} focus on the single-input-single-output (SISO) network whose objective is the signal-to-noise ratio (SNR)---which is directly related to the signal power, whereas this work focuses on the MIMO case with a much more complicated objective that not only depends on the power but also depends on the channel matrix rank. Second, the SISO blind beamforming can be interpreted as aligning each reflected channel with the direct channel vector, whereas the MIMO case generalizes each channel to a matrix and thus the aligning interpretation no longer applies. Third, the performance analysis (i.e., how many random samples are needed) is different from the SISO case \cite{ren2022configuring}.

\emph{Notation:} For a complex number $u$, $\mathfrak{Re}\{u\}$ is the real part, and $u^*$ is the complex conjugate. For a matrix $\bA$, $[\bA]_{ij}$ is the $(i,j)$th entry, $\det(\bA)$ is the determinant, $\|\bA\|_F$ is the Frobenius norm, and $\bA^\hh$ is the conjugate transpose. Let $\bI_n$ be the $n\times n$ identity matrix. Moreover, write $f(n)=\Omega(g(n))$ if there is $c>0$ such that $f(n) \geq c g(n)$ for $n$ sufficiently large.


\section{System Model}
\label{sec:sys}

Consider an IS-assisted MIMO transmission with $M$ transmit antennas and $L$ receive antennas. Assume that the IS has $N$ REs. Denote by $\Theta=(\theta_1, \theta_2,\ldots,\theta_N)$ the phase shift array of the IS where $\theta_n$ is the phase shift of the $n$th RE. Assume also that each phase shift $\theta_n$ is limited to a discrete set 
\begin{align}
    \Phi_K=\{0, \omega, 2\omega, \ldots, (K-1)\omega \} \ \ \text{where} \ \omega=\frac{2\pi}{K},
\end{align}
for some positive integer $K\ge 2$.
Denote by $\bD \in \mathbb C^{L\times M}$ the channel from the transmitter to the receiver, denote by $\bF \in \mathbb C^{N \times M}$ the channel from the transmitter to the IS, and denote by $\bG \in \mathbb C^{L \times N}$ the channels from the IS to the receiver. More specifically, the $(i,j)$th entry of $\bD$ is the channel from the $j$th transmit antenna to the $i$th receive antenna, the $(n,j)$th entry of $\bF$ is the channel from the $j$th transmit antenna to the $n$th RE, and the $(i,n)$th entry of $\bG$ is the channel from the $n$th RE to the $i$th receive antenna. With the diagonal matrix $\bm\Psi=\mathrm{diag}(e^{j\theta_1}, e^{j\theta_2},\ldots,e^{j\theta_N})$, the overall channel $\bH\in\mathbb C^{L\times M}$ from transmitter to receiver is given by
\begin{equation}
    \bH = \bD +\bG\bm\Psi\bF.
\end{equation}
Let $\bW\in\mathbb{C}^{M\times S}$ be the beamforming matrix at the transmitter, where $S\leq \min\{M, L\}$ is the number of data streams. The transmitter has the power constraint $\|\bW\|^2_F\le P$. Let $\sigma^2$ be the background noise power. With the background complex Gaussian noise $\bm {z} \sim\mathcal{CN}(0,\sigma^2 \bI_L)$ and the transmit signal $\bs\sim\mathcal{CN}(0, \bI_S)$, the received signal $\by\in\mathbb C^{L}$ is given by
\begin{equation}
    \by = \bH\bW\bs + \bm{z},
\end{equation}
and the channel capacity is given by
\begin{equation}
C =\log\det\bigg(\bI_L+\frac{1}{\sigma^2}\bH\bW\bW^\hh\bH^\hh\bigg). 
\end{equation}
To achieve the maximum channel capacity, we consider the joint active and passive beamforming problem as
\begin{subequations}
\label{discrete problem}
\begin{align}
    \underset{\bW,\,\Theta}{\text{maximize}} &\quad \log\det\bigg(\bI_L+\frac{1}{\sigma^2}\bH\bW\bW^\hh\bH^\hh\bigg)\\
    \text {subject to} &\quad \theta_n\in \Phi_K,\;n=1,2,\ldots,N, \\
    &\quad \|\bW\|^2_F\leq P.
\end{align}
\end{subequations}
We propose optimizing two variables $\bW$ and $\Theta$ alternatingly (e.g., as in \cite{MIMO_Zhangsw} and \cite{zhao2022TCOM}). In particular, when $\Theta$ is held fixed, the optimization problem with respect to $\bW$ is the traditional MIMO beamforming problem in \cite{tse2005wireless}, so the rest of this paper focuses on optimizing $\Theta$ in problem \eqref{discrete problem} with $\bW$ being fixed.

\begin{figure}
    \centering
    \includegraphics[width=0.5\linewidth]{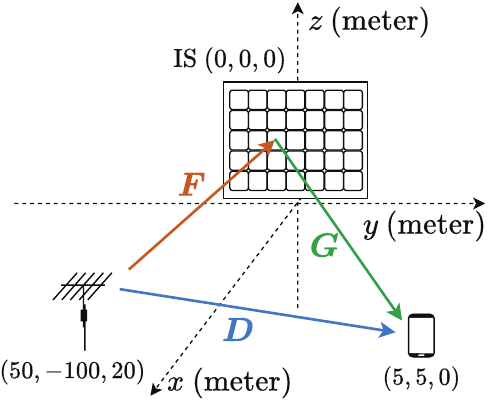}
    \caption{An example of the IS-assisted MIMO system.}
    \label{fig:simu_system}
\end{figure}

For fixed $\bW$, we define a new matrix $\bA=\bH\bW$; the MIMO channel capacity can then be reformulated as
\begin{align}
\label{eq:def_capacity}
C &=\ \log\det\bigg(\bI_L+\frac{1}{\sigma^2}\bA\bA^\hh\bigg).
\end{align}
For the positive semi-definite matrix $\bA\bA^\hh$, sort its eigenvalues as $\lambda_1\ge\lambda_2\ge\ldots\ge\lambda_S\ge0$.
The channel capacity in \eqref{eq:def_capacity} can be further rewritten as
\begin{equation}
\label{eqn:capacity}
    C = \sum^S_{i=1}\log\bigg(1+\frac{\lambda_i}{\sigma^2}\bigg).
\end{equation}
As such, problem \eqref{discrete problem} with respect to the passive beamforming $\Theta$ can finally be formulated as
\begin{subequations}
\label{IRS_passive_problem}
\begin{align}
    \underset{\,\Theta}{\text{maximize}} &\quad \sum^S_{i=1}\log\bigg(1+\frac{\lambda_i}{\sigma^2}\bigg)\\
    \text {subject to} &\quad \theta_n\in \Phi_K,\;n=1,2,\ldots,N.
\end{align}
\end{subequations}
We remark that CSI is not available in the above problem. Note that the optimization of $\bW$ still requires the CSI, but it only depends on the overall channel superposition from transmitter to receiver, so it does not require the CSI of each reflected channel. Thus, optimizing $\bW$ requires low-dimensional CSI, while optimizing $\Theta$ does not require any CSI.

\section{Passive Beamforming with CSI}
\label{sec:CSI_known}

Our ultimate goal is to solve the problem \eqref{IRS_passive_problem} without any CSI, but let us first assume that the CSI is known \emph{a priori}. In this section, we approximate the capacity maximization problem \eqref{IRS_passive_problem} as a sum power maximization problem, thereby developing a linear-search algorithm for passive beamforming.

We start by showing that the capacity maximization problem \eqref{IRS_passive_problem} can be approximated as a sum power maximization problem. Although \cite{MIMO_Zhangsw} introduces a similar approximation, our paper gives a new insight: the approximation problem can be obtained from both the upper bound and the lower bound.

Considering the lower approximation, we show that
\begin{equation}
    \label{eq:capacity_lower_bound}
    C\stackrel{(a)}{\geq} \log\Bigg(1+\frac{\max_{i}\{\lambda_i\}}{\sigma^2} \Bigg)\stackrel{(b)}{\geq}  \log\Bigg(1+\frac{1}{\sigma^2 S}\sum^S_{i=1}\lambda_i \Bigg),
\end{equation}
where $(a)$ follows as we only keep the largest $\log(1+\lambda_i/\sigma^2)$ while deleting the rest from \eqref{eqn:capacity}, and $(b)$ follows as $\log(1+x)$ is a non-decreasing function and we raise $x$ from $\frac{1}{\sigma^2 S}\sum^S_{i=1}\lambda_i$ to $\max_{i}\{\lambda_i\}/\sigma^2$.
In the meanwhile, by Jensen's inequality, we have the upper approximation
\begin{equation}
    \label{eq:capacity_upper_bound}
    C \leq S\log\Bigg(1+\frac{1}{\sigma^2S}\sum^S_{i=1}\lambda_i \Bigg).
\end{equation}
Thus, when $S$ is held fixed, whichever bound is used to approximate $C$, problem \eqref{IRS_passive_problem} can be always converted to
\begin{align}
    \label{sum_eigenvalue_problem}\underset{\Theta\in\Phi^N_K}{\text{maximize}}&\quad \sum^S_{i=1} \lambda_i,
\end{align}
where  $\Phi^N_K=\Phi_K\times\Phi_K\times\cdots\times\Phi_K$ is a Cartesian product of $N$ sets of $\Phi_K$. Notice that problem \eqref{sum_eigenvalue_problem} is equivalent to the sum power maximization problem
\begin{align}
\label{prob sum power}
    \underset{\Theta\in\Phi^N_K}{\text{maximize}}&\quad f(\Theta):=\sum^L_{i=1} \sum^S_{j=1} |[\bA]_{ij}|^2,
\end{align}
due to the basic properties of the Frobenius norm: $\sum_{i=1}^S\lambda_i = \|\bm A\|_F^2 = \sum^L_{i=1} \sum^S_{j=1} |[\bA]_{ij}|^2$. We from now on consider the above new problem and denote its objective function by $f(\Theta)$.

\begin{figure}[t]
\small
\vspace{-10pt}
\begin{algorithm}[H]
\caption{Blind Beamforming for MIMO Transmission}
\label{alg:SPCSM}
\begin{algorithmic}[1]
    \State\textbf{input:} $\Phi_K$ and $N$.
    \For{$t=1,2,\ldots,T$}
    \State Generate each $\theta_{nt}$ uniformly from $\Phi_K$.
    \State Measure the received signal powers $\left|y_{it}\right|^2$ under $\Theta_t$.
    \EndFor
    \State Compute $\widehat{\mathbb{E}}\left[g\mid \theta_n=k \omega\right]$ as in \eqref{eq:MIMO_CSM}.
    \State Decide each $\theta_n$  as in \eqref{eq:SPCSM_solution}.
    \State \textbf{Output:} Phase shift array $\Theta^{\mathrm{Blind}}$.
\end{algorithmic}
\end{algorithm} 
\end{figure}

Let $\alpha_{ij}=[\bD\bW]_{ij}$ and $\beta_{inj}=[\bG]_{in}\cdot[\bF\bW]_{nj}$.
Then, we show that the sum channel power with respect to each receive antenna $i$ can be lower bounded as
\begin{align}
\label{eq:lower_bound_AS}
\sum^S_{j=1} |[\bA]_{ij}|^2 &\geq 4\cdot\sum^S_{j=1}\sum^N_{n=1}\mathfrak{Re}\Big\{\alpha^*_{ij} \beta_{inj}e^{j\theta_n}\Big\},
\end{align}
which follows by the arithmetic-geometric mean inequality
\begin{equation}
\left|\alpha_{ij}\right|^2+\left|\sum^N_{n=1}\beta_{inj}e^{j\theta_n}\right|^2\geq 2\cdot\sum^N_{n=1}\mathfrak{Re}\{\alpha^*_{ij} \beta_{inj}e^{j\theta_n}\}.\notag
\end{equation}
Furthermore, substituting the inequality \eqref{eq:lower_bound_AS} into $f(\Theta)=\sum^L_{i=1} \sum^S_{j=1} |[\bA]_{ij}|^2$, we bound $f(\Theta)$ in \eqref{prob sum power} from below as
\begin{equation}
f(\Theta)\ge f_b(\Theta):=4\cdot\sum^L_{i=1}\sum^S_{j=1}\sum^N_{n=1}\mathfrak{Re}\Big\{\alpha^*_{ij} \beta_{inj}e^{j\theta_n}\Big\}.
\end{equation}
Notice that if we use the above lower bound $f_b(\Theta)$ to approximate $f(\Theta)$, each phase shift can be optimally determined by linear search:
\begin{equation}
\label{eq:LS_solution}
    \theta_n^{\mathrm{CSI}} = \arg\max_{\varphi\in\Phi_K}\sum^L_{i=1} \sum^S_{j=1} \mathfrak{Re}\Big\{\alpha^*_{ij} \beta_{inj}e^{j\varphi}\Big\}.
\end{equation}
With every $\theta_n$ optimized in this manner, the overall complexity equals $O(NSL)$. Apart from the low complexity, the greater advantage of the linear search method is that it can be implemented without CSI, as shown in the following section.

\section{Blind Passive Beamforming without CSI}

Recall that our ultimate goal is to solve the passive beamforming problem without CSI. This section proposes a blind beamforming algorithm that optimizes $\Theta$ in the absence of CSI. The main result of this section is to show that the CSI-based solution in \eqref{eq:LS_solution} can be recovered in the absence of CSI, namely blind beamforming.

The main idea of the proposed blind beamforming algorithm is to randomly configure the IS and then decide each $\theta_n$ based on the CSM of the received signal power. We first generate and try out $T$ random samples of $\Theta$ with each $\theta_{n}$ drawn from $\Phi_K$ uniformly and independently, for $n=1,\ldots, N$ and $t=1,\ldots,T$. Let $\theta_{nt}$ be the realization of $\theta_n$ in the $t$th random sample, and accordingly let $\Theta_t=(\theta_{1t}, \theta_{2t}, \ldots, \theta_{Nt})$.
For each $\Theta_t$, we measure the corresponding received signal power at each receive antenna; denote by $|y_{it}|^2$ the received signal power at the $i$th receive antenna for the $t$th random sample. 
Next, we group the $T$ random samples with respect to each $n=1,\ldots,N$ and each $k=1,\ldots,K$ as:
\begin{equation}
    \mathcal{Q}_{n k} =\left\{t \mid \theta_{nt}=k\omega\right\}.
\end{equation}
i.e., $\mathcal{Q}_{n k}$ is a set of sample indices in which the phase shift of the $n$th RE equals $k\omega$.

For each random sample $\Theta_t$, compute 
\begin{equation}
\label{eq:MIMO utility}
g(\Theta_t)=\sum_{i=1}^L |y_{it}|^2.
\end{equation}
Based on $g(\Theta_t)$, we compute the CSM \cite{ren2022configuring} as
\begin{align}
    \label{eq:MIMO_CSM}
    \widehat{\mathbb E}\left[ g\mid\theta_{n}=k\omega\right]=\frac{1}{|\mathcal{Q}_{n k}|} \sum\limits_{t \in \mathcal{Q}_{n k}} \sum_{i=1}^L |y_{it}|^2.
\end{align}
Finally, we decide each phase shift $\theta_n$ by comparing the CSM:
\begin{align}
\label{eq:SPCSM_solution}
    \theta_n^{\mathrm{Blind}} = \arg \max_{\varphi\in\Phi_K} \widehat{\mathbb E}\left[ g\mid\theta_{n}=\varphi\right].
\end{align}
The above process is summarized in Algorithm \ref{alg:SPCSM}. We remark that the proposed algorithm reduces to the existing algorithm in \cite{ren2022configuring} when the number of transmit antennas $M=1$.

We now provide performance analysis. As stated in the following proposition, the blind beamforming in Algorithm \ref{alg:SPCSM} can recover the linear search method in \eqref{eq:LS_solution}.
\begin{proposition}
\label{prop:equivalent}
The blind beamforming solution $\Theta^{\mathrm{Blind}}$ in \eqref{eq:SPCSM_solution} converges to the linear-search solution $\Theta^{\mathrm{CSI}}$ in \eqref{eq:LS_solution} in probability so long as $ T = \Omega(LN^2S^2\left(\log(NL)\right)^3)$.
\end{proposition}
\begin{IEEEproof}
    See Appendix.
\end{IEEEproof}

\section{Field Tests in Real-World 5G Networks}

\begin{figure*}[t]
\centering
\includegraphics[width=0.8\linewidth]{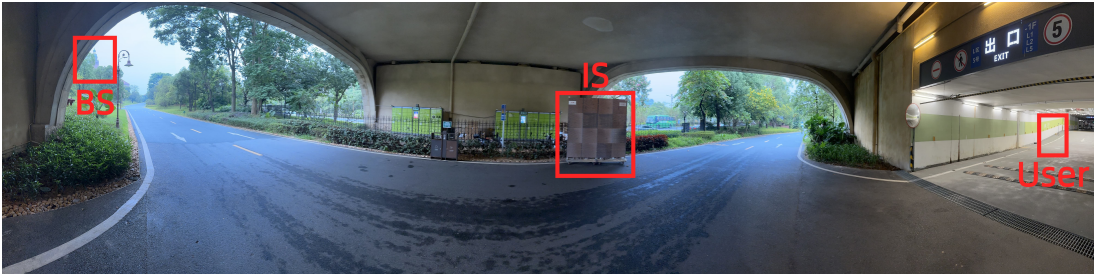}
\caption{A panoramic view of the parking lot. The BS is located on a 20-meter-high terrace. The user terminal is located inside the entrance of the underground parking lot. The IS machine-I is placed at the entrance, approximately 250 meters from the BS and 25 meters from the user terminal.}
\label{fig:indoor_environment}
\end{figure*}

\begin{figure}[t]
    \centering
    \includegraphics[width=3.5cm]{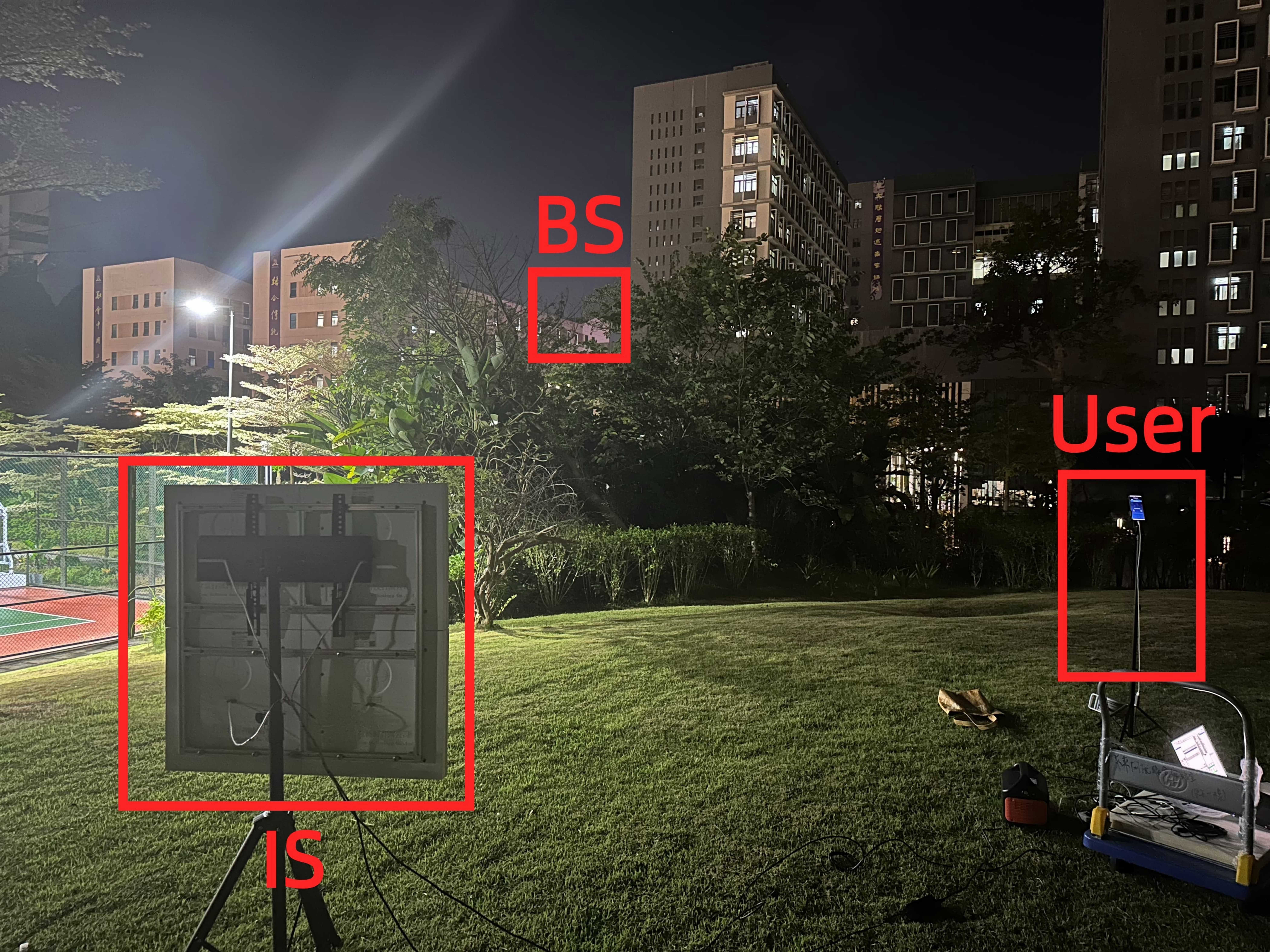}
    \caption{The scene of the outdoor testbed. The BS is deployed on the roof of a 6-storey building. The BS-to-IS distance is about 300 meters, while the IS-to-user distance is about 8 meters.}
	\label{fig:outdoor_environment}
\end{figure}

We carry out the field tests in both indoor and outdoor scenarios, as shown in Fig.~\ref{fig:indoor_environment} and Fig.~\ref{fig:outdoor_environment}, to evaluate the performance of the proposed blind beamforming algorithm. The public base station (BS) is equipped with 64 transmit antennas and the user terminal is equipped with four receive antennas. The carrier frequency is $2.6$ GHz or $3.5$ GHz and the bandwidth of signals is $200$ MHz. We remark that throughout the field tests, we cannot control the BS, so no additional knowledge about the BS is known. Two types of IS are used:
\begin{itemize}
    \item \emph{IS machine-I:} It comprises $256$ REs and each RE provides $4$ phase shift options $\{0,\frac{\pi}{2},\pi\, \frac{3\pi}{2}\}$, i.e., $N_1=256$ and $K=4$. The operating frequency is $2.6$ GHz.
    \item \emph{IS machine-II:} It comprises $400$ REs and each RE provides $4$ phase shift options $\{0,\frac{\pi}{2},\pi\, \frac{3\pi}{2}\}$, i.e., $N_2=400$ and $K=4$. The operating frequency is $3.5$ GHz.
\end{itemize}

We compare the following methods in the field tests:
\begin{itemize}
    \item \emph{Zero Phase Shift (ZPS):} Fix all phase shifts to be zero.
    \item \emph{Beam Training \cite{mei2021mIRSMIMO,2023HuanghTVT}:} Try out $T$ random samples of $\Theta$ and choose the best in terms of the summed received power at the receiver.
    \item \emph{Rank Beam Training:} Try out $T$ random samples of $\Theta$ and choose the best in terms of the channel rank.
    \item \emph{Rank CSM:} Take the channel rank as the utility of the general CSM method proposed in \cite{ren2022configuring}.
\end{itemize}
Let $T=1000$. We consider the reference signal received power (RSRP) gain and the data rate gain as compared to the baseline without using the IS.

\begin{table}
\footnotesize
\renewcommand{\arraystretch}{1.3}
\centering
\caption{\small Field Test Results}
\begin{tabular}{|l|rr|rr|}
\firsthline
\multicolumn{1}{|c|}{}       & \multicolumn{2}{c|}{RSRP Gain (dB)}                           & \multicolumn{2}{c|}{Rate Gain (Mbps)}                            \\ \hline
\multicolumn{1}{|l|}{Method} & \multicolumn{1}{r}{Indoor} & \multicolumn{1}{r|}{Ourdoor} & \multicolumn{1}{r}{Indoor} & \multicolumn{1}{r|}{Ourdoor} \\ \hline
ZPS    & 6.10 & 0.72 & 21.07 & 8.94   \\
Beam Training   & 6.21 & 0.92 & 21.72 & 12.82\\
Rank Beam Training  & 6.47 & 0.80 & 22.22 & 21.35\\
Rank CSM  & 5.05 & 0.52 & 5.58  & 8.19 \\
Proposed Algorithm & 9.92 & 2.47 & 49.88  & 26.14 \\
\lasthline
\end{tabular}
\label{tab:field_test_results}
\end{table}


Table \ref{tab:field_test_results} summarizes the RSRP boost and data rate boost of different algorithms in both the indoor and outdoor tests. As shown in Table \ref{tab:field_test_results}, the proposed algorithm can significantly increase RSRP and data rate and outperform the benchmarks in the indoor scenario. For instance, it can increase the transmission rate by about $50$ Mbps, which improves upon the best of the remaining algorithms by about $124\%$. When it comes to the outdoor scenario, it can be seen that the proposed algorithm still performs the best. However, the superiority is not as significant as that in the indoor scenario, which may be because of the small size of the IS machine-II and the severe pathloss at $3.5$ GHz frequency.



\section{Simulations}

\begin{figure*}[t]
\centering
    \begin{minipage}[t]{0.3\linewidth}
    \centering
    \includegraphics[width=1.1\linewidth]{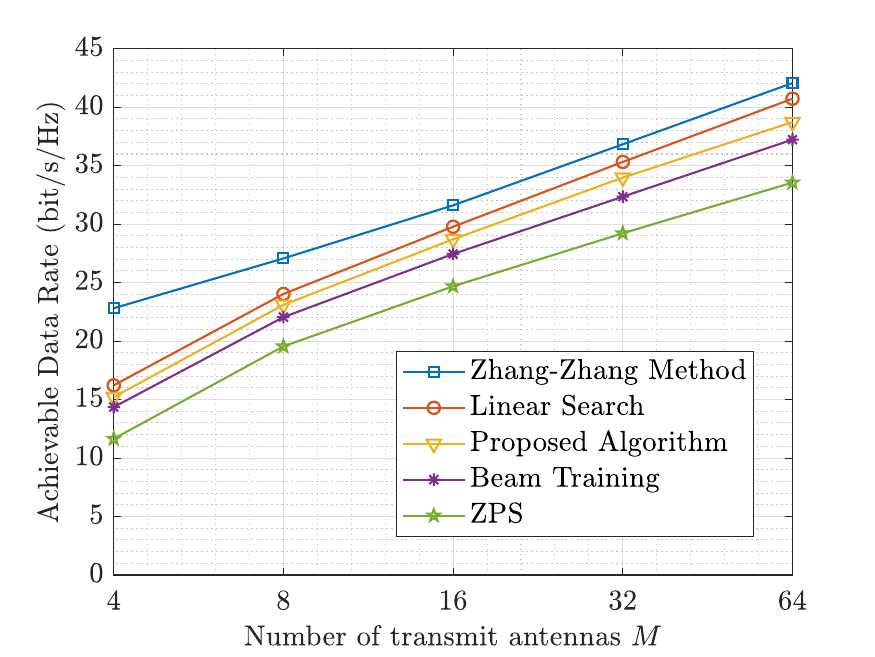}
    \centering
    \caption{Rate vs. number of transmit antennas.}
    \label{fig:rate_vs_m}
    \end{minipage} 
    \quad
    \begin{minipage}[t]{0.3\linewidth}
        \centering
    \includegraphics[width=1.1\linewidth]{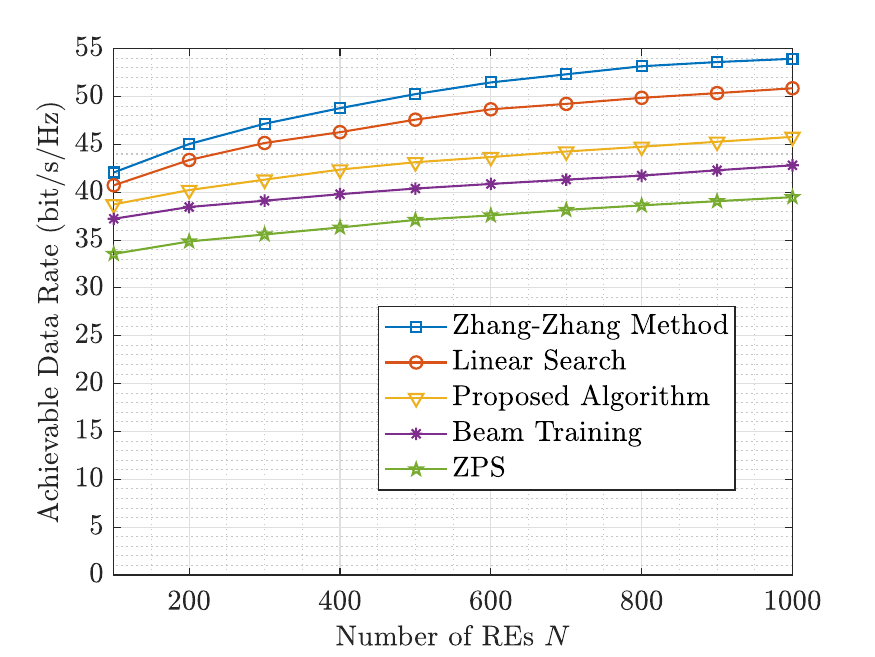}
    \caption{Rate vs. number of REs.}
    \label{fig:rate_vs_n}
    \end{minipage}
    \quad
    \begin{minipage}[t]{0.3\linewidth}
        \centering
    \includegraphics[width=1.1\linewidth]{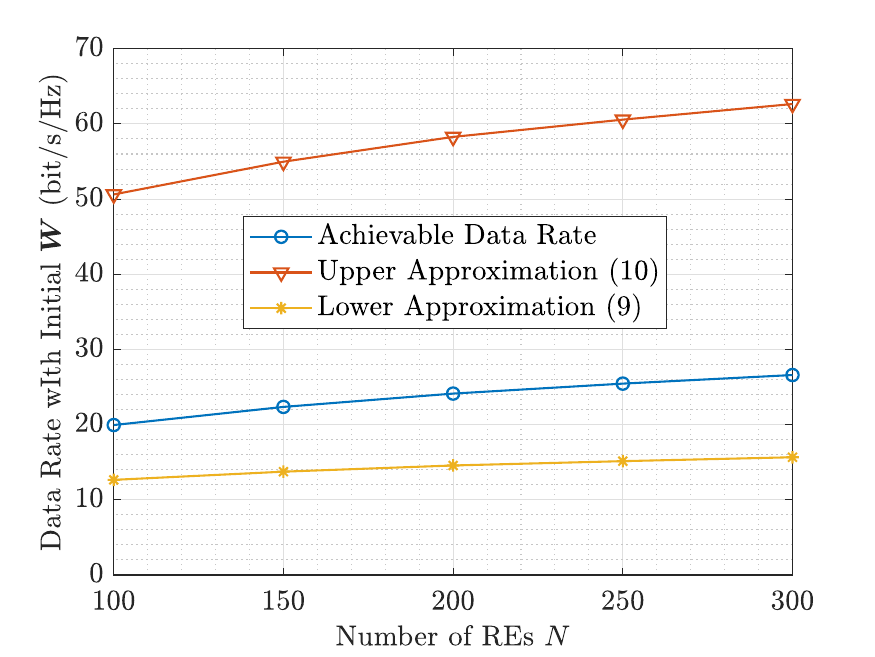}
    \caption{Approximation gap vs. number of REs.}
    \label{fig:gap}
    \end{minipage}
\end{figure*}


We further compare the proposed methods with the existing CSI-based methods through simulations as shown in Fig.~\ref{fig:simu_system}. The transmit antennas of the BS and REs of the IS are arranged as a half-wavelength
spaced uniform linear array (ULA); the carrier frequency equals 2.6 GHz, so the wavelength $\lambda\approx10$ cm and thus the
transmit antenna and RE spacing equals 5 cm. Our pathloss model follows \cite{ren2022configuring, Xu2024Coordinating}, which are generated as $\text{PL}_\text{BU}=32.6+36.7\log_{10} d_\text{BU}$ (in dB), $\text{PL}_\text{BI}=30+22\log_{10} d_\text{BI}$ (in dB), and $ \text{PL}_\text{IU}=30+22\log_{10} d_\text{IU}$ (in dB),
where $\text{PL}_\text{BU}$, $\text{PL}_\text{BI}$ and $\text{PL}_\text{IU}$ are the pathloss factors from BS to user, from BS to IS, and from IS to user, respectively, while $d_\text{BU}$, $d_\text{BI}$ and $d_\text{IU}$ are the corresponding distance in meters. Our channel model follows the existing works \cite{Wang2023MIMO,pan2020multicell}. In particular, the direct channel $\bD$ is modeled as the Rayleigh fading channel while the reflected channels $\bF$ and $\bG$ are modeled as the Rician fading channels. The transmit power constraint is $30$ dBm, and the background noise level is $-90$ dBm. Let $K=4$ by default. In the simulation, the number of random samples is $T=1000$.

Aside from the ZPS and beam training methods compared in the field tests, the \emph{Zhang-Zhang Method} proposed in \cite{MIMO_Zhangsw} is also considered as a benchmark. Note that the \emph{Zhang-Zhang Method} considers the continuous beamforming; we round the obtained continuous solution to the discrete set. Moreover, the computational complexities of the proposed blind beamforming, the linear search method, the beam training, and the \emph{Zhang-Zhang Method} are $O(N(T+K))$, $O(NK)$, $O(T)$, and $O(LMNI)$, respectively, where $I$ is the number of iterations required by the \emph{Zhang-Zhang Method} to converge.

Fig.~\ref{fig:rate_vs_m} shows the achievable data rate versus the number of transmit antennas $M$ when the number of receive antennas and the number of REs are fixed at $L=4$ and $N=100$, respectively. Observe that the data rates achieved by the different algorithms all increase with $M$. Observe also that the proposed algorithm can further improve upon beam training by more than 1 bit/s/Hz. Fig.~\ref{fig:rate_vs_n} further shows the achievable rate versus the number of REs $N$ when the number of transmit antennas and receive antennas are fixed at $M=64$ and $L=4$, respectively. Note that the gap between the \emph{Zhang-Zhang} method \cite{MIMO_Zhangsw} and the proposed algorithm grows with $N$ at the beginning and then converges when $N\ge 600$. A plausible explanation follows. When $N$ is small (e.g., when $N=100$), the system capacity is limited, so the further performance gain obtained from the CSI knowledge as in the Zhang-Zhang method is limited as well. With $N$ increasing, the performance gap becomes larger and ultimately converges. Nevertheless, the proposed blind beamforming is still much preferable for two reasons. First, the \emph{Zhang-Zhang Method} requires full CSI, but the proposed method does not require any CSI to optimize $\Theta$. Second, the computation burden of the blind beamforming is much smaller, as shown in Table \ref{tab:run_time}. Fig.~\ref{fig:gap} shows the gap between the data rate and its upper/lower bounds in \eqref{eq:capacity_lower_bound} and \eqref{eq:capacity_upper_bound} when different numbers of REs are considered, with $\bW$ fixed at the starting point. We find that the gap becomes larger when the number of REs is increased; this makes sense since the approximation error grows with the problem size.

\begin{table}[t]
\footnotesize
    \renewcommand{\arraystretch}{1.3}
\centering
\caption{\small Running Time of Different Methods When $M=64, L=4$.}
\begin{tabular}{lrrrr}
\firsthline
& \multicolumn{3}{c}{Running Time (second)}\\
\cline{2-4}
Method      & $N=100$ & $N=200$ & $N=300$\\
\hline
Zhang-Zhang Method  & 1.84 & 6.56 & 15.06 \\
Linear Search    & 0.013  & 0.025 & 0.042 \\
Proposed Algorithm  & 0.62  & 1.27 & 2.23 \\
Beam Training & 0.29  & 0.59 & 1.26 \\
\lasthline
\end{tabular}
\label{tab:run_time}
\end{table}

\section{Conclusion}
\label{sec:Conclusion}
This paper aims to maximize the channel capacity of an IS-aided point-to-point MIMO system with the challenging discrete phase shift constraint. We show that the MIMO channel capacity maximization problem can be approximated as a more tractable sum power maximization problem, thus enabling the implementation of a search-based method with linear complexity. Furthermore, we reveal that the search-based method can be implemented even without CSI. Both field tests and simulations demonstrate the superior performance of the proposed blind beamforming scheme.

\appendix
We would let $\widehat{\mathbb E}\left[ g\mid\theta_{n}=k\omega\right]$ be approximately the same as $\mathbb E\left[ g\mid\theta_{n}=k\omega\right]$ so that $\theta_n^{\mathrm{Blind}}=\theta_n^{\mathrm{CSI}}$. To this end, we aim at the condition that
$\left|\widehat{\mathbb{E}}[|y_i|^2\mid\theta_n=k\omega]-\mathbb{E}[|y_i|^2\mid\theta_n=k\omega]\right|<\epsilon_{in}$ for each tuple $(i, n, k)$, where $\epsilon_{in}>0$ is the difference between the largest and the second largest values of $\sum^S_{j=1}\mathfrak{Re}\Big\{Q_{ij}^\hh R_{inj}e^{jk\omega}\Big\}$ when $k$ varies. 


For a particular tuple $(i, n, k)$, define the error event
\begin{equation}
\mathcal{E}_{ink}=\left\{\left|\widehat{\mathbb{E}}\left[|y_i|^2 \mid \theta_n=k \omega\right]-\mathbb{E}\left[|y_i|^2 \mid \theta_n=k \omega\right]\right| \geq \epsilon_0\right\} \notag,
\end{equation}
where $\epsilon_0=\inf_{(i,n)}\left\{ \epsilon_{in}\right\}$. Following the steps in \cite{ren2022configuring}, we can bound the error probability as
\begin{align}
& \mathbb{P}\left\{\mathcal{E}_{in k}\right\} \leq 2 e^{-\frac{2 \epsilon_0^2  T}{9 q^2 \nu^2K}}+\frac{(9 \sigma^4+18 q \nu \sigma^2) K}{\epsilon_0^2 T}+8e^{-\frac{q}{2}},
\end{align}
where $\nu=\inf_i\left\{\sum_{j=1}^S|Q_{ij}|^2+\sum_{n=1}^N\sum_{j=1}^S|R_{inj}|^2\right\}$ and $q > 0$ is an arbitrary positive constant.

Now consider the overall error event $\mathcal{E}_0=$ $\bigcup_{(i, n, k)} \mathcal{E}_{in k}$, the probability of which can be bounded by the union bound as
\begin{equation}
\begin{aligned}
    \mathbb{P}\left\{\mathcal{E}_0\right\}
    & \leq 2NKL e^{-\frac{2 \epsilon_0^2  T}{9 q^2 \nu^2K}}+\frac{9\sigma^4 NK^2L}{\epsilon_0^2 T}\notag\\
    &\qquad\qquad+\frac{18 q \nu \sigma^2NK^2L}{ \epsilon_0^2 T} +8 N KL e^{-q / 2}.
\end{aligned}
\end{equation}
Further, for any $0<\xi<1$ and any $0<p_0 \leq \xi/4$, we have
\begin{align*}
8 N KL e^{-\frac{q}{2}} \leq p_0\;&\text{if }q\geq 2\log(8 N KL/p_0),\\
\frac{9 \sigma^4 N K^2 L}{\epsilon_0^2 T}\leq p_0\;&\text{if } T\geq \frac{9 \sigma^4 N K^2 L}{\epsilon_0^2 p_0},\\
\frac{18 q\nu \sigma^2 N K^2L}{ \epsilon_0^2 T}\leq p_0\;&\text{if }T\geq\frac{18 q\nu \sigma^2 N K^2L}{ \epsilon_0^2 p_0},\\
2 N KL e^{-\frac{2 \epsilon_0^2  T}{9 q^2 \nu^2  K}}\leq p_0,\;&\text{if }T\geq \frac{9 q^2 \nu^2  K\log(2NKL/p_0)}{2 \epsilon_0^2}.
\end{align*}
Putting the above results together, we show that $\mathbb{P}\{\mathcal{E}_0\}\leq4p_0\le\xi$ when $T = \Omega(L\nu^2(\log(NL))^3)$.
Since $\nu\propto NS$, we ultimately have $\mathbb{P}\{\mathcal{E}_0\}\leq\xi$ and thus $\mathbb{P}\left\{\Theta^{\mathrm{Blind}}=\Theta^{\mathrm{CSI}}\right\}\geq 1-\xi$ for an arbitrarily small $\xi>0$, provided that $T =\Omega(LN^2S^2(\log(NL))^3)$.

\bibliographystyle{IEEEtran}
\bibliography{IEEEabrv,Ref}
\end{document}